\documentclass[12pt,preprint]{aastex}
\usepackage{natbib}

\begin{document}
 
\newcommand{\kms}{km~s$^{-1}$}
\newcommand{\cm}[1]{\, {\rm cm^{#1}}}
\newcommand{\N}[1]{{N({\rm #1})}}
\newcommand{\Lya}{Ly$\alpha$}
\newcommand{\lya}{Ly$\alpha$}
\newcommand{\delv}{\Delta v}

\title{Elemental Abundances in Two High Column Density 
Damped Lyman Alpha Systems at $z<1.5$ \altaffilmark{1}}  

\altaffiltext{1}{Based on data obtained with the 
NASA/ESA {\it Hubble Space Telescope} operated by STScI-AURA under NASA 
contract NAS 5-26555.}

\author{Sandhya M. Rao\altaffilmark{2}, 
Jason X. Prochaska\altaffilmark{3,4},
J. Christopher  Howk\altaffilmark{4,5}, 
and Arthur M. Wolfe\altaffilmark{5}}
                                                                             
\altaffiltext{2}{Department of Physics \& Astronomy, University of Pittsburgh,
Pittsburgh, PA 15260; rao@everest.phyast.pitt.edu}

\altaffiltext{3}{UCO/Lick Observatory, University of California, Santa
Cruz, CA 95064}

\altaffiltext{4}{Department of Physics, and Center for Astrophysics and
Space Sciences, University of California, San Diego, C-0424, La Jolla, CA
92093}
\altaffiltext{5}{Visiting Astronomer, W. M. Keck Telescope.
The Keck Observatory is a joint facility of the University
of California and the California Institute of Technology.}

\begin{abstract}

We present Keck/HIRES abundance measurements and metal-line kinematic
profiles of the damped Lyman alpha systems (DLAs) towards the  quasars
Q0933+733 ($z_{abs}=1.479$) and Q0948+433 ($z_{abs}=1.233$).  These
two DLAs have among the five highest \ion{H}{1} column densities
at any redshift: $N(HI)=4.2\times10^{21}$ cm$^{-2}$.
The metal-line data, presented here for the first time,
reveal that these DLAs are noteworthy for several other reasons as
well.  
1) The Q0933+733 DLA exhibits simple kinematic structure with
unusually narrow velocity widths as measured from its unsaturated
metal lines  ($\delv=16$ \kms). At 2.6\% solar, it has the second
lowest  metallicity at $z<2$. 
2) The Q0948+433 DLA has among the
strongest metal-line transitions of any known DLA. The saturated
\ion{Si}{2} $\lambda$1808 line implies a high metallicity ([Si/H]$> -1$)
and a significant $\alpha$-enhancement. The  
strong metal lines of this DLA  have made possible the
detection of \ion{Ti}{2} $\lambda$1910,   \ion{Co}{2} $\lambda$2012,
and \ion{Mg}{1} $\lambda$2026.  
3) We find that the relative gas-phase
abundances of both DLAs follow the general trend seen at high
redshift, e.g., enhanced Zn/Fe and Si/Fe,  and sub-solar Mn/Fe,
indicating that  there is little evolution in the nucleosynthetic
patterns of DLAs down to this epoch.  
4) Their high \ion{H}{1} column
densities imply that these DLAs  dominate the column density-weighted
cosmic mean metallicity, $\left<Z\right>$, of the universe at $z<1.5$.  
Using the 15 DLAs with measured metallicities in the redshift
interval $0.4<z<1.5$,  we find $\left<Z\right> = -0.89^{+0.40}_{-0.33}$,
where the uncertainties are 95\% confidence limits. 

\end{abstract}

                                                                               
\section{Introduction}

It has been known for almost two decades now that damped Lyman alpha
systems (DLAs) seen in the spectra of background quasars trace the
neutral gas content of the universe back to the redshifts of the most
distant quasars (Wolfe et al. 1986). This fact has since been
exploited to study the evolution of the neutral gas content of the
universe (e.g., Lanzetta et al. 1991, Peroux et al. 2003, Rao \&
Turnshek 2000 [RT00], Storrie-Lombardi \& Wolfe 2000; Prochaska \&
Herbert-Fort 2004)  and its chemical enrichment history (e.g., Pettini
et al. 1999,  Prochaska et al.\ 2003b, and references therein). Until
recently, however, the rarity of DLAs in combination  with the
requirement of the {\it Hubble Space Telescope} to discover them at
redshifts $z<1.65$ has meant that most of our knowledge about the
galaxies that contain the  bulk of the neutral gas has come from
objects that span only the first 30\% of the age of  the universe.

High resolution spectrometers on 10m-class telescopes have been
instrumental  in advancing our knowledge about the detailed abundance
patterns and kinematic properties of DLAs (e.g.\ Pettini et al. 1999;
Pettini et al. 2000;  Molaro et al.\ 2000; Prochaska et al. 2001;
Ledoux, Bergeron, \& Petitjean 2002; Prochaska \& Wolfe 2002;
Dessauges-Zavadsky et al. 2004).  This work has mainly concentrated on
the large sample of $z>1.65$ DLAs known to date.  With the success of
UV surveys for DLAs at redshifts $z<1.65$ (RT00; Rao, Turnshek, \&
Nestor  2004), the number of DLAs available for study at these
redshifts has been steadily increasing and it is now becoming possible
to fill in the large gap in cosmic time between today  and $z \sim 2$.
The importance of studying low-redshift DLAs cannot be understated:
these systems bridge the gap between our knowledge of  present-day
galaxies and high-redshift DLAs.  The low-redshift systems are the
only  population (currently) of cosmological objects for which the
neutral gas content, abundances, and kinematic properties obtained via
high resolution spectroscopy  can be compared with observable galaxy
properties such as morphology and star formation history.

To date,  only a handful of $z<1.5$ DLAs  have been observed  with
high resolution spectroscopy \citep[e.g.][]{ptt99,ptt00}.  These
observations  showed that the abundance ratios and general kinematic
properties of lower-redshift DLAs are similar to those at higher
redshift.  Here we add to the growing list of DLA abundance
measurements at $z<1.65$ with Keck/HIRES data on the DLAs towards the
quasars Q0933+733 ($z_{abs} = 1.479$) and Q0948+433 ($z_{abs} =
1.233$).  With $N(HI)=4.2\times10^{21}$ cm$^{-2}$ (\S2.1), these two
DLAs rank among the five highest \ion{H}{1} column density systems
known.  High column density DLAs are important for follow-up
metallicity observations for two reasons.  First, their high column
density may allow the  measurement of weak lines rarely observed in
DLAs. For example, we report a $>3\sigma$ detection of \ion{Co}{2}
$\lambda 2012$ in the Q0948+433, $z=1.233$, DLA (\S3.2).  This is the
second detection of Co to date \citep{ellison01} and the most secure.
Its detection promises the observation of weak transitions of even
rarer elements (e.g.,\ Cu, Ga, B, etc.).  Second, the highest column
density DLAs dominate the determination of cosmic metallicity since
cosmic metallicity is given by an \ion{H}{1} column density-weighted
mean.  Thus, the current measurement of the cosmological mean
metallicity in neutral gas at $z\approx 1.4$ is dominated by these two
DLAs.  This paper is organized as follows. In \S2 we describe the
observations and data reduction techniques. The metal-line profiles
and chemical abundances  of the two DLAs are presented in \S3. We
conclude with a discussion in \S4.

\section{Observations}

\subsection{{\it HST} data}

\subsubsection{Q0933+733}
{\it HST}-UV spectroscopic data for Q0933+733 were obtained as part of
a Cycle 6 survey for DLAs in known $z<1.65$ \ion{Mg}{2}
absorption-line systems ({\it HST} PID 6577, S. Rao PI; RT00). This
$z_{em}=2.525$ quasar has two strong \ion{Mg}{2} systems along the
sightline at $z=1.4789$ and $z=1.4973$ (Steidel \& Sargent 1992).
Figure 10 of RT00 shows the {\it HST}-FOS G270H grating spectrum of
Q0933+733. Since the Ly$\alpha$ profiles of the two absorbers are
blended with each other and with a Lyman limit edge, the \ion H1
column density of the DLA at $z=1.479$ was determined using the
Ly$\beta$ absorption line (see figure 11 of RT00), and was found  to
be $(4.2\pm0.8) \times 10^{21}$ cm$^{-2}$. The higher redshift system,
which also has detectable Ly$\beta$ absorption, has an H\,I column
density $N(HI)=(1.0\pm0.5) \times 10^{20}$ cm$^{-2}$ (Rao et al.\
2004) just below the DLA criterion, and we reserve an  analysis of
this absorber to a future paper.  The uncertainty in these two column
density measurements reflects the low signal-to-noise ratio of the
spectrum near the Ly$\beta$ lines.

\subsubsection{Q0948+433}
This quasar was observed with {\it HST}-STIS as part of another DLA
survey for strong \ion{Mg}{2} absorption-line systems found in optical
spectra of quasars ({\it HST} PID 9051, R. Becker PI). The data, which
we obtained from the {\it HST} public  archive, consist of two
exposures taken with the G230L grating  for a total of 2,333
seconds. We combined the two sub-exposures using a weighted average
based on exposure time and combined the error arrays accordingly. We
then normalized the resulting spectrum using standard routines in
IRAF.  Since there are Lyman limit edges in this spectrum, only a
local continuum near  the DLA line was determined.  Part of the
normalized {\it HST} archive spectrum encompassing the DLA line is
shown in Figure 1. As is usually the case  for high column density DLA
lines, the Ly$\alpha$ forest populates the DLA trough making it very
difficult to use a routine such as least squares minimization to fit a
Voigt profile to the data. Therefore, the best `fit'  was estimated by
eye. A Voigt profile with column density $N(HI)=4.2 \times 10^{21}$
cm$^{-2}$ convolved with the Gaussian line spread  function of the
{\it HST}-STIS G230L grating  is shown as the dotted line in Figure 1.
The uncertainty in the column density determination, $0.5\times
10^{21}$ cm$^{-2}$ in this case, is dominated by the error in
continuum placement and was estimated using the procedure described in
RT00.

\subsection{Keck/HIRES data}
 
Q0933+733 and Q0948+433 were observed with Keck/HIRES  \citep{vogt94}
on UT 2003 March 11 and 12  for a total integration time of 5,400s
each.  For both sets of observations, we implemented the C1 decker
(0.86$''$ width; FWHM~$\approx 6$  \kms\ resolution) and the kv380
filter to block second order light.  Data reduction and calibration
proceeded in standard fashion \citep[e.g.][]{pro01} with the MAKEE
reduction package kindly provided by T. Barlow.  The data were
continuum fit with custom software and rebinned to  2 \kms\ per pixel.
The signal-to-noise ratio of these data exceeds 20 per pixel for the
majority of the Q0933+733 spectrum and is greater than $10$ per pixel
redward of 4500\AA\ in the Q0948+433 spectrum.

\section{Metal-Line Profiles and Chemical Abundances}

All of the ionic column densities were derived with the apparent
optical  depth method  \citep[AODM;][]{sav91,jenkins96}.   This
technique can be used to identify  hidden saturation by  comparing the
apparent column densities of multiple transitions from a single ion.
This technique also gives an efficient, non-parametric  means of
calculating total column densities  for each ion.  For those
transitions where the profile saturates in at least one pixel (i.e.,
normalized flux $<0.05$), the column densities are listed as lower
limits.   We report non-detections as $3\sigma$ statistical upper
limits.  We have ignored continuum error in our analysis which may
dominate the measurements of very weak transitions.  We estimate a
systematic error $< 10\%$ in most cases due to continuum placement.
We calculate ionic column densities for each transition and assign
final column densities by calculating the weighted-mean for ions with
multiple transitions.  In the velocity plots, $v=0$ is chosen
arbitrarily and corresponds to the redshift listed in the figure
caption.  We indicate regions of blending, primarily through blends
with other metal-line systems or the \lya\ forest, by plotting with
dotted lines.

The atomic data considered in this paper are listed in
Table~\ref{tab:fosc}. Table~\ref{tab:solabd} gives our assumed solar
abundance data  compiled by \cite{grvss99}.

\subsection{Q0933+733; z=1.479}

The metal line profiles of this  $z=1.479$ DLA are shown  in
Figure~\ref{fig:0933mtl}.  The profiles are remarkably narrow with a
velocity width $\delv = 16$ \kms, where $\delv$ is  the  interval
encompassing 90\% of the total optical depth \citep{pw97}, and  is
measured using the unsaturated Zn\,II 2026 profile.   This is among
the lowest $\delv$ value recorded for a DLA \citep{pw01}. We also note
a likely  'edge-leading asymmetry' for the profiles towards positive
velocities in  the unsaturated metal lines, consistent with rotation.
This characterization, however, is limited by the fact that the
profile is only $\approx 2$ resolution elements wide.

We have integrated the line profiles using the AODM to calculate the
ionic column densities given in Table~\ref{tab:Q0933+732_1.479}.
Columns 1 and 2 list the ion and rest wavelength, column 3 gives  the
column density, column 4 is the weighted mean column density
determined from multiple absorption lines of the same ion, and is
listed alongside the first occurrence of the ion, and column 5 is the
derived abundance relative to solar.  Comparisons of multiple
transitions from a given ion (e.g.\ \ion{Cr}{2}) are in reasonable
agreement and there is no indication of line saturation.  We note,
however, that the \ion{Cr}{2} $\lambda$2062 and  \ion{Zn}{2}
$\lambda$2062 column densities are lower than their partner
transitions which may indicate that the continuum level is
systematically low in this region.

Measurements of Zn and Si  indicate a metallicity $\sim 1/40$ solar
which  falls $\approx 0.5$\,dex below the mean metallicity at $z \sim
2$.  The value lies at the lower end of the DLA metallicity
distribution at these redshifts and is the second lowest value
recorded at $z<2$.  Figure~\ref{fig:0933rel} describes the relative
abundances for this DLA system. As is standard practice in  stellar
abundance studies, and now generally adopted to describe DLA
abundances, we plot measured abundances relative to Fe (top
panel). This  comparison serves to illustrate the combined effects of
dust depletion and nucleosynthesis \citep[e.g.][]{pw02}. The bottom
panel gives the absolute abundances on a logarithmic scale where
hydrogen has value 12.0.  In both cases we present gas-phase
abundances, i.e.,\ the observed values uncorrected for depletion.

Similar to the results for other $z \sim 1.5$ DLAs \citep{ptt99} and
the majority of $z>2$ DLAs \citep{pw02}, this DLA shows modest
enhancements of Zn and Si relative to Fe.  These enhancements are
suggestive of differential depletion and/or nucleosynthetic processes,
and it is difficult to  disentangle the two effects.  The Cr/Fe
overabundance relative to solar is notable as it exceeds the value
observed for Cr/Fe in the majority of Milky Way sightlines
\citep[e.g.][]{sav96}.  \cite{pw02} reported a similar enhancement for
their sample of high $z$ DLAs and suggested that it is a signature of
nucleosynthesis.   Interestingly, the enhancement runs contrary to the
behavior of Cr in  metal-poor halo stars where [Cr/Fe] decreases with
[Fe/H] \citep{mcw95b}.  Although this is only a modest departure from
the solar ratio, it suggests differences in the yields of Fe-peak
elements.

Finally, we note sub-solar Ti/Fe and Mn/Fe ratios.  As emphasized by
\cite{dessauges02}, Mn and Ti are particularly valuable for
disentangling the competing effects of nucleosynthesis and
differential depletion.  The sub-solar Ti/Fe ratio observed here is
indicative of differential depletion, and is somewhat surprising given
the low dust depletion levels implied by the relatively low Zn/Fe
ratio.   Perhaps the low Ti/Fe ratio is more reflective of grain  core
growth as opposed to grain mantle growth and, in turn, an important
clue to the mechanism of grain formation  in young, metal-poor
galaxies (e.g., supernovae, red giant winds).  In contrast to Ti, the
observed low Mn/Fe ratio is a reflection of nucleosynthesis.  The
sub-solar value follows previous measurements of Mn/Fe
\citep{ptt00,ledoux02} and supports the notion that Mn production is
metallicity-dependent \citep{mcwilliam03}.

We note that except for \ion{Ti}{2} and \ion{Fe}{2}, the results of
Khare et al. (2004) agree with our measurements for this DLA. Their
Multiple  Mirror Telescope ({\it MMT}) spectra do not have the high
resolution and signal-to-noise ratio per pixel that characterize our
Keck data.  Thus, we believe that the \ion{Ti}{2} column density is
indeed an upper limit as shown in Table 3, rather than a confirmed
detection as claimed by Khare et al. (2004). In addition, our Fe
abundance is more secure since  Khare et al. (2004) used the
equivalent widths of lines measured by Steidel \& Sargent (1992) and
an assumed {\it b} parameter to estimate its column density.

\subsection{Q0948+433; z=1.233}

The metal line profiles of the damped \lya\ system at $z=1.233$ are
shown  in Figure~\ref{fig:0948mtl}.   The line profiles show greater
complexity than the DLA toward Q0933+733 although the observed
velocity  width ($\delv = 68$ \kms) is still lower than the median
value at $z>2$.  We also note that the line profiles show the
`edge-leading' characteristic of many high $z$ DLAs \citep{pw97} and a
mild asymmetry to the line profiles.  Again, these characteristics are
consistent with rotation.

The metal-line transitions of this DLA are among the strongest ever
observed.  Note, in particular, the saturated \ion{Si}{2}
$\lambda$1808 profile which implies $\N{SiII} > 10^{16} \cm{-2}$ and
[Si/H]~$> -1$\,dex.  This \ion{Si}{2} $\lambda$1808 profile is deeper
than that observed for almost any sightline through the Milky
Way\footnote{Due to depletion of Si from the gas-phase in  Milky Way
sightlines with large $\N{HI}$ and because sight lines observed in the
Milky Way at high resolution tend to have lower $\N{HI}$.}. We find,
therefore, that  this `metal-strong' DLA shows positive detections of
weak transitions like \ion{Ti}{2} $\lambda$1910,  \ion{Co}{2}
$\lambda$2012, and \ion{Mg}{1} $\lambda$2026. We note that for the
\ion{Ti}{2} $\lambda$1910 doublet, we integrated the optical depth
over both transitions and used the sum of the oscillator strengths to
convert that optical depth to column density.  The column densities
for all of the detected transitions are listed in
Table~\ref{tab:Q0948+43_1.233}.  We predict that the \ion{Cr}{2} 2026
transition contributes $<5\%$ to the \ion{Zn}{2} 2026 profile and have
ignored this blend in our analysis.

The relative abundances of this DLA are presented in
Figure~\ref{fig:0948rel}.  The abundance pattern of Cr, Mn, Zn, and Fe
is similar to that of the DLA system toward Q0933+733: Cr/Fe and Zn/Fe
show modest enhancements and Mn/Fe is significantly underabundant.  In
addition, we report the second detection of Co to date.  In contrast
to the first detection of Co by Ellison et al. (2001) in the
Q2206$-$199 DLA system\footnote{But see Prochaska \& Wolfe (2002)  who
do not consider this detection to be very secure.}, which is
overabundant  relative to Fe, we find a solar Co/Fe abundance ratio in
the  Q0948+433 DLA. Ellison et al. show that the overabundance in the
Q2206$-$199 DLA is consistent with the pattern seen in Galactic bulge
stars. The solar Co/Fe ratio and observed metallicity (e.g., [Fe/H])
in the Q0948+433 DLA, on the other hand,  are more typical of
metal-rich Galactic halo stars (Gratton \& Sneden 1991).  While it is
impossible to establish trends with only two systems, it is  at least
evident that their [Co/Fe] ratios fall within the locus traced by
Galactic stars.

Perhaps the most interesting result of these observations is that both
Si and Ti\footnote{One may view the Ti result cautiously because of
the relatively poor S/N at the Ti\,II 1910 transitions.}  (the
$\alpha$-elements) are significantly enhanced relative to Fe.   Since
Ti is more readily adsorbed onto dust grains than Fe
\citep[e.g.][]{sav96}, differential depletion leads to  negative
values for [Ti/Fe]. Therefore, the observed enhancement   (i.e.,
positive [Ti/Fe])  is not due to depletion, but implies a
nucleosynthetic origin instead.  This characteristic is, as in the
case of the Co/Fe ratio discussed above,  typical of Galactic halo
stars \citep[e.g.,][]{gratton91}.  If one were to correct for
differential depletion in this DLA, by using the Cr/Zn ratio for
example, the dust-corrected Ti/Fe ratio may exceed +0.4\,dex.
Differential depletion could be invoked to explain the large Si/Fe
ratio, but we emphasize that [Si/Zn]~$> +0.1$, and given the
significant line saturation in  \ion{Si}{2}$\lambda$1808, the Si/Zn
ratio may exceed +0.3\,dex.  Altogether, these relative abundances
indicate a significant $\alpha$-enhancement in this modest metallicity
gas.  This result contradicts a primary conclusion of  \cite{ptt00}
who argued that the low $z$ DLA systems have  solar $\alpha$/Fe ratios
and therefore may not trace the bulk of star formation.  We note that
the abundance pattern and metallicity mirror those observed for the
most metal-rich Galactic halo stars \citep[e.g.][]{edv93} and the
Galactic thick disk stars \citep{pro00}.  Perhaps we are observing the
by-products of the generation of star formation which immediately
proceeds the formation of the galactic disk in this system, a
speculation supported by the 'edge-leading' nature  of the metal-line
profiles.

\section{Discussion}

The DLAs towards Q0933+733 and Q0948+433 are unusual in several
ways. At $4.2\times10^{21}$ cm$^{-2}$, their \ion{H}{1} column
densities are among the five highest at any redshift of over 150  DLAs
with measured metallicities, and so they dominate the column
density-weighted mean metallicity of the universe at $z\approx 1.4$.
Here we update the  Prochaska et al. (2003b) estimate of the
column-density weighted  mean metallicity for the redshift interval
$0.4<z<1.5$. We add  the Q0933+733 DLA to the sample, and use a
revised  value for the  \ion{H}{1} column density of the Q0948+433
DLA.  We also include the recent results of Turnshek et al. (2004) and
Khare et al. (2004). Table 5 lists the data used to determine the
column density-weighted mean metallicity,  $\left<Z\right>=\log
[\Sigma_i 10^{[\rm M/\rm H]_i} N(HI)_i/\Sigma_i N(HI)_i]$.  We find
$\left<Z\right>=-0.89^{+0.40}_{-0.33}$ at a median redshift of
$\left<z\right>=0.86$, where the uncertainties have been
estimated using the bootstrap technique and are 95\% confidence
limits. This result is shown in  Figure 6 along with high-redshift
measurements from Prochaska et al. (2003b).  The new low-redshift 
value of $\left<Z\right>$ is in agreement with the  value estimated 
by Prochaska et al. (2003b), but with significantly smaller uncertainty. 
We also confirm the trend of increasing metallicity with cosmic
time and, from a least-squares fit to the  $\left<Z\right>$ values,
derive a slope of $m=-0.26 \pm 0.06$.

The narrow velocity width of the Q0933+733 DLA, $\delv = 16$ \kms,
implies a kinematically simple system, while the edge-leading
asymmetries of the unsaturated lines reveal structure even on these
small velocity scales. Although the metal line profiles of the
Q0948+433 DLA have several components, their combined velocity width
of $\delv = 68$ \kms\ is still lower than the median value measured
for DLAs. Both DLAs follow the  trend first noted by \cite{wol98},
which is still evident in the larger sample of \cite{pro03a},  that
high $N(HI)$ DLAs tend to have smaller metal-line velocity widths.
The compilation of 21 cm line widths by \cite{kc03} indicates that a
similar trend might be present between 21 cm line widths and $N(HI)$.
Additionally, \cite{Nestor03} found a correlation between DLA
metallicity  and \ion{Mg}{2} rest equivalent width, the latter being
an indicator of the  velocity spread along the line of sight. Thus, a
comprehensive study of 21 cm absorption line properties, \ion{H}{1}
column densities, metallicities, and   metal line  profiles for DLAs
in front of radio loud quasars might provide some insight  into the
nature of the individual cloud(s) that gives rise to a DLA system.

The high $N(HI)$ value of the Q0933+733 DLA ensures that many
metal-line  transitions are easily measured even at a metallicity as
low as 2.6\% solar (\S3.1). This is the second lowest value recorded
at $z<2$. Even so,  relative abundance patterns in this DLA are
similar to results for other DLAs, implying that there is minimal
evolution in the  nucleosynthetic patterns for gas with [M/H]~$<-1$ at
this epoch.

Finally, the Q0948+433 DLA is characterized by the highly saturated
\ion{Si}{2}$\lambda$1808 line, which implies a high metallicity
([Si/H]~$>-1$) and a significant $\alpha$-enhancement.  The unusually
strong lines of this DLA have made possible the detections  of
\ion{Ti}{2} $\lambda$1910,   \ion{Co}{2} $\lambda$2012, and
\ion{Mg}{1} $\lambda$2026, and we further predict that echelle
observations to the atmospheric limit would reveal positive detections
of transitions for Ga, P, Ge, Sn, and possibly Pb.  Since the mean
metallicity of the DLAs is observed to increase with decreasing
redshift, a large sample of $z<1.5$ DLAs may be more likely to yield
`metal-strong' candidates.  Therefore, surveys of these galaxies may
be the most efficient approach to obtain comprehensive abundance
pattern measurements of young galaxies, and given the recent loss of
STIS on HST, emphasizes the need for a UV spectrograph in space.

\acknowledgments

S.R. gratefully acknowledges Dave Turnshek for many helpful
discussions.  The space-based component of this work has been funded
in part by  NASA-LTSA grant NAG5-7930.  The authors wish to recognize
and acknowledge the very significant cultural role and reverence that
the summit of Mauna Kea has always had within the indigenous Hawaiian
community.  We are most fortunate to have the opportunity to conduct
observations from this mountain.  We acknowledge the Keck support
staff for their efforts in performing these observations.  J.X.P. and
A.M.W. are partially supported by NSF grant AST0307824.

\clearpage

\begin{deluxetable}{lcccc}
\tablewidth{0pc}
\tablecaption{ATOMIC DATA \label{tab:fosc}}
\tabletypesize{\footnotesize}
\tablehead{\colhead{Transition} &\colhead{$\lambda$} &\colhead{$f$} & \colhead{Ref}}
\startdata
   CI 1656 & 1656.9283 & 0.1405 &  1  \\
 AlII 1670 & 1670.7874 & 1.8800 &  1  \\
 PbII 1682 & 1682.150  & 0.15636&  1  \\
 NiII 1741 & 1741.5531 & 0.0427 &  7  \\
 NiII 1751 & 1751.9157 & 0.0277 &  7  \\
 SiII 1808 & 1808.0130 & 0.00218600 & 11  \\
AlIII 1854 & 1854.7164 & 0.539 &  1  \\
AlIII 1862 & 1862.7895 & 0.268 &  1  \\
TiII 1910a & 1910.6000 & 0.202 & 12  \\
TiII 1910b & 1910.9380 & 0.098 & 12  \\
CoII 2012  & 2012.1664 & 0.03679 &  8  \\
 ZnII 2026 & 2026.1360 & 0.489 & 13  \\
CrII 2026  & 2026.269  & 0.00471 & 15 \\
MgI 2026   & 2026.4768 &  0.1120 & 1 \\
 CrII 2056 & 2056.2539 & 0.105 & 13  \\
 CrII 2062 & 2062.2340 & 0.078 & 13  \\
 ZnII 2062 & 2062.6640 & 0.256 & 13  \\
 CrII 2066 & 2066.1610 & 0.0515 & 13  \\
 FeII 2249 & 2249.8768 & 0.00182100 & 14  \\
 FeII 2260 & 2260.7805 & 0.00244 & 14  \\
 FeII 2344 & 2344.2140 & 0.114 &  3  \\
 FeII 2374 & 2374.4612 & 0.0313 &  3  \\
 FeII 2382 & 2382.7650 & 0.320 &  3  \\
 MnII 2576 & 2576.8770 & 0.3508 &  1  \\
 FeII 2586 & 2586.6500 & 0.0691 &  3  \\
 MnII 2594 & 2594.4990 & 0.271 &  1  \\
 FeII 2600 & 2600.1729 & 0.239 &  3  \\
 MnII 2606 & 2606.462  & 0.1927&  1 \\
 MgII 2796 & 2796.3520 & 0.6123 & 15  \\
 MgII 2803 & 2803.5310 & 0.3054 & 15  \\
  MgI 2852 & 2852.9642 & 1.810 &  1  \\
\enddata
\tablerefs{1:
\cite{morton91}; 3: \cite{morton04}; 
7: \cite{fedchak00}; 
8: \cite{mullman98};
11: \cite{bergs93b}; 
12: \cite{wiese01}; 13: \cite{bergs93}; 14: \cite{bergs94}; 15:
\cite{verner94}}
\end{deluxetable}

\newpage
\begin{deluxetable}{rrr}
\tablewidth{0pc}
\tablecaption{ADOPTED SOLAR ABUNDANCES\label{tab:solabd}}
\tabletypesize{\footnotesize}
\tablehead{\colhead{Elm} &\colhead{$\epsilon(X)$} & \colhead{Z}}
\startdata
H & 12.00 &  1\\
B &  2.79 &  5\\
Mg&  7.58 & 12\\
Al&  6.49 & 13\\
Si&  7.56 & 14\\
S &  7.20 & 16\\
Ti&  4.94 & 22\\
Cr&  5.67 & 24\\
Mn&  5.53 & 25\\
Fe&  7.50 & 26\\
Co&  4.91 & 27\\
Ni&  6.25 & 28\\
Zn&  4.67 & 30\\
\enddata
\end{deluxetable}
\clearpage

\begin{deluxetable}{lcccc}
\tablewidth{0pc}
\tablecaption{IONIC COLUMN DENSITIES: Q0933+733, $z = 1.479$ \label{tab:Q0933+732_1.479}}
\tabletypesize{\footnotesize}
\tablehead{\colhead{Ion} &\colhead{$\lambda$} &\colhead{AODM$^a$} & \colhead{$N_{\rm adopt}$}& \colhead{[X/H]}}
\startdata
H I & 1215.7&& $21.62^{+0.10}_{-0.10  }$ \\
C  I  &1656.9&$<12.737$\\  
Al II &1670.8&$>13.367$&$>13.367$&$>-2.743$\\  
Al III&1854.7&$12.718 \pm  0.036$\\  
Al III&1862.8&$12.472 \pm  0.094$\\  
Si II &1808.0&$>15.555$&$>15.555$&$>-1.625$\\  
Ti II &1910.6&$<12.433$&$<12.433$&$<-2.127$\\  
Cr II &2056.3&$13.584 \pm  0.016$&$13.563 \pm  0.012$&$-1.727 \pm  0.101$\\  
Cr II &2062.2&$13.510 \pm  0.023$\\  
Cr II &2066.2&$13.606 \pm  0.029$\\  
Mn II &2576.9&$12.962 \pm  0.013$&$12.957 \pm  0.011$&$-2.193 \pm  0.101$\\  
Mn II &2606.5&$12.947 \pm  0.018$\\  
Fe II &2249.9&$15.178 \pm  0.016$&$15.190 \pm  0.011$&$-1.930 \pm  0.101$\\  
Fe II &2260.8&$15.203 \pm  0.016$\\  
Fe II &2344.2&$>14.511$\\  
Fe II &2374.5&$>14.882$\\  
Fe II &2382.8&$>14.110$\\  
Fe II &2586.7&$>14.649$\\  
Fe II &2600.2&$>14.185$\\  
Co II &2012.2&$<12.969$&$<12.969$&$<-1.561$\\  
Ni II &1741.6&$13.916 \pm  0.014$&$13.922 \pm  0.011$&$-1.948 \pm  0.101$\\  
Ni II &1751.9&$13.936 \pm  0.020$\\  
Zn II &2026.1&$12.748 \pm  0.019$&$12.712 \pm  0.019$&$-1.578 \pm  0.102$\\  
Zn II &2062.7&$12.551 \pm  0.058$\\  
Pb II &1682.2&$<12.320$&$<12.320$&$< 0.640$\\  
\enddata
\tablenotetext{a}{Errors reflect statistical uncertainty.  One should adopt an 
additional 15$\%$ systematic error for weak transitions due to continuum uncertainty.}
\end{deluxetable}

\clearpage 
 
\begin{deluxetable}{lcccc}
\tablewidth{0pc}
\tablecaption{IONIC COLUMN DENSITIES: Q0948+433, $z = 1.233$ \label{tab:Q0948+43_1.233}}
\tabletypesize{\footnotesize}
\tablehead{\colhead{Ion} &\colhead{$\lambda$} &\colhead{AODM$^a$} & \colhead{$N_{\rm adopt}$}& \colhead{[X/H]}}
\startdata
H I & 1215.7&& $21.62^{+0.05}_{-0.06  }$ \\
Mg I  &2026.5&$13.069 \pm  0.044$\\  
Mg I  &2853.0&$>12.878$\\  
Mg II &2796.4&$>14.036$&$>14.306$&$>-2.894$\\  
Mg II &2803.5&$>14.306$\\  
Al III&1854.7&$13.111 \pm  0.045$\\  
Al III&1862.8&$13.119 \pm  0.088$\\  
Si II &1808.0&$>16.147$&$>16.147$&$>-1.033$\\  
Ti II &1910.6&$13.339 \pm  0.070$&$13.339 \pm  0.070$&$-1.221 \pm  0.092$\\  
Cr II &2056.3&$13.912 \pm  0.016$&$13.905 \pm  0.014$&$-1.385 \pm  0.062$\\  
Cr II &2066.2&$13.889 \pm  0.025$\\  
Mn II &2576.9&$13.368 \pm  0.015$&$13.305 \pm  0.012$&$-1.845 \pm  0.061$\\  
Mn II &2594.5&$13.249 \pm  0.025$\\  
Mn II &2606.5&$13.253 \pm  0.024$\\  
Fe II &2249.9&$15.536 \pm  0.019$&$15.563 \pm  0.013$&$-1.557 \pm  0.061$\\  
Fe II &2260.8&$15.595 \pm  0.018$\\  
Fe II &2344.2&$>14.676$\\  
Fe II &2374.5&$>15.209$\\  
Fe II &2382.8&$>14.296$\\  
Fe II &2586.7&$>14.870$\\  
Co II &2012.2&$13.025 \pm  0.119$&$13.025 \pm  0.119$&$-1.505 \pm  0.133$\\  
Zn II &2026.1&$13.152 \pm  0.013$&$13.152 \pm  0.013$&$-1.138 \pm  0.061$\\  
\enddata
\tablenotetext{a}{Errors reflect statistical uncertainty.  One should adopt an 
additional 15$\%$ systematic error for weak transitions due to continuum uncertainty.}
\end{deluxetable}

\newpage

\begin{deluxetable}{lccccc}
\tablewidth{0pc}
\tablecaption{$0.4<z<1.5$ Sample of DLA Metallicities}
\tabletypesize{\footnotesize}
\tablehead{\colhead{QSO} &\colhead{$z_{abs}$} &\colhead{$\log N(HI)$} & \colhead{[M/H]}& 
$f_{[\rm M/\rm H]}$\tablenotemark{a} & \colhead{Refs.}}
\startdata
Q1229$-$021  & 0.395 & $20.75 \pm 0.07$ & $-0.47 \pm 0.15$ & 1 & a    \\
Q0235+164    & 0.526 & $21.65 \pm 0.04$ & $-0.58 \pm 0.15$ & 2 & b    \\  
Q1622+238    & 0.656 & $20.36 \pm 0.10$ & $-0.87 \pm 0.25$ & 3 & c, d \\
Q1122$-$168  & 0.682 & $20.45 \pm 0.05$ & $-1.00 \pm 0.15$ & 3 & e    \\
Q1328+307    & 0.692 & $21.25 \pm 0.06$ & $-1.20 \pm 0.09$ & 1 & f    \\
Q1323-0021   & 0.716 & $20.54 \pm 0.15$ & $ 0.04 \pm 0.16$ & 1 & g,h  \\
Q1107+0048   & 0.740 & $21.00 \pm 0.05$ & $-0.64 \pm 0.07$ & 1 & g,h  \\
Q0454+039    & 0.860 & $20.69 \pm 0.06$ & $-0.79 \pm 0.12$ & 4 & i    \\
Q1727+5302   & 0.945 & $21.16 \pm 0.04$ & $-0.58 \pm 0.15$ & 1 & j    \\
\nodata      & 1.031 & $21.41 \pm 0.04$ & $-1.32 \pm 0.38$ & 1 & j    \\
Q0302$-$223  & 1.009 & $20.36 \pm 0.11$ & $-0.73 \pm 0.12$ & 4 & i    \\
Q0948+433    & 1.233 & $21.62 \pm 0.05$ & $-1.14 \pm 0.06$ & 1 & k    \\
Q0935+417    & 1.373 & $20.52 \pm 0.10$ & $-0.94 \pm 0.13$ & 1 & f    \\
Q1354+258    & 1.420 & $21.54 \pm 0.06$ & $-1.61 \pm 0.16$ & 1 & i    \\ 
Q0933+733    & 1.479 & $21.62 \pm 0.10$ & $-1.58 \pm 0.10$ & 1 & k    \\
\enddata

\tablenotetext{a}{ 1: Zn;
2: Xray measurement; range includes assumed solar and 
$\alpha$-enhanced metal ratios in absorber with a solar O abundance of 8.74, 
and  Galactic gas with ISM metallicity (see Turnshek et al. 2003);
3: Fe + 0.4;
4: Si, S, or O.
}
\tablerefs{
a. Boiss\'e et al. 1998;
b. Turnshek et al. 2003; 
c. Churchill et al. 2000;
d. RT00;
e. de la Varga et al. 2000;
f. Meyer et al. 1995;
g. Rao, Turnshek, \& Nestor 2004 [N(HI) measurements];
h. Khare et al. 2004 [N(ZnII) measurements];
i. Pettini et al. 2000;
j. Turnshek et al. 2004;
k. This work.
}
\end{deluxetable}

\clearpage

\begin{figure}
\plotone{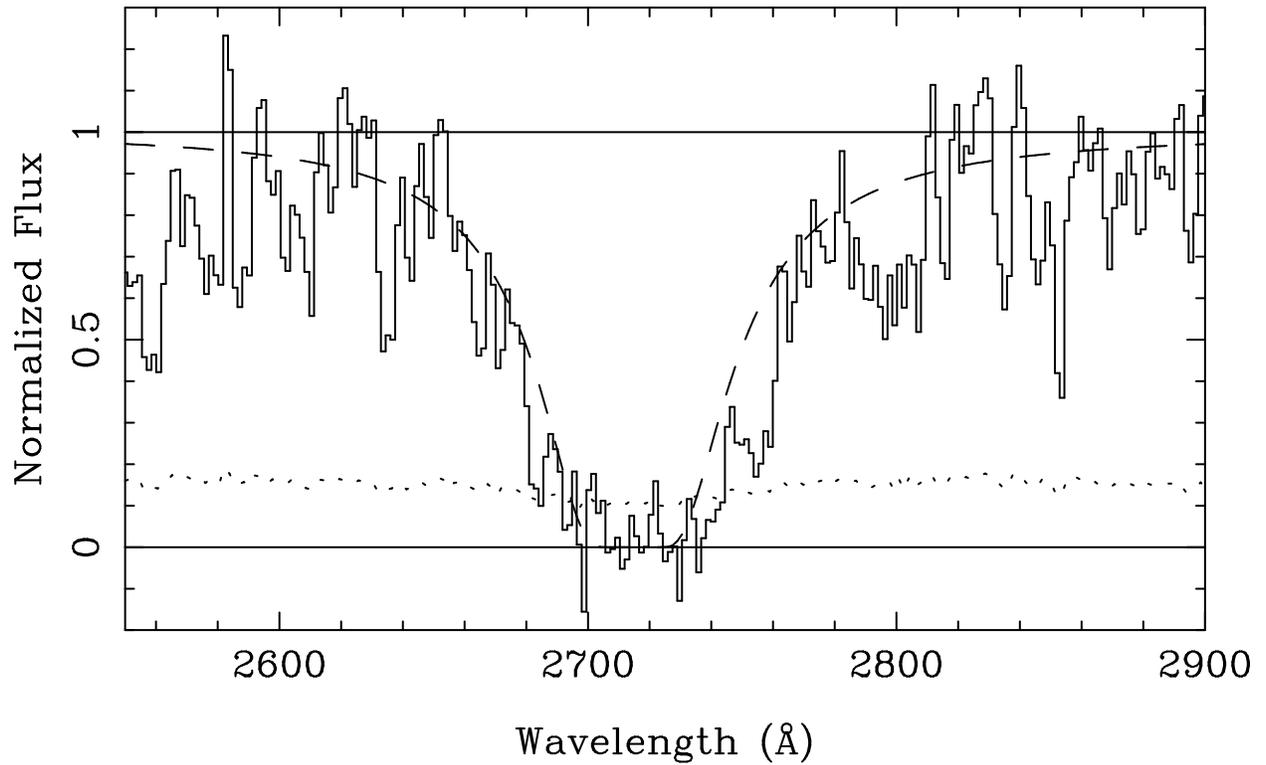}
\caption{Part of the {\it HST}-STIS G230L spectrum of Q0948+433 showing 
the DLA line at 2715\AA\ ($z_{abs}=1.233$). The dashed line is a Voigt
profile convolved with the line spread function of the {\it HST}-STIS
G230L grating and has column density $N(HI)=4.2 \times 10^{21}$ cm$^{-2}$.
The 1$\sigma$ error array is shown as a dotted line.  
}
\end{figure}

\begin{figure}
\epsscale{0.85}
\plotone{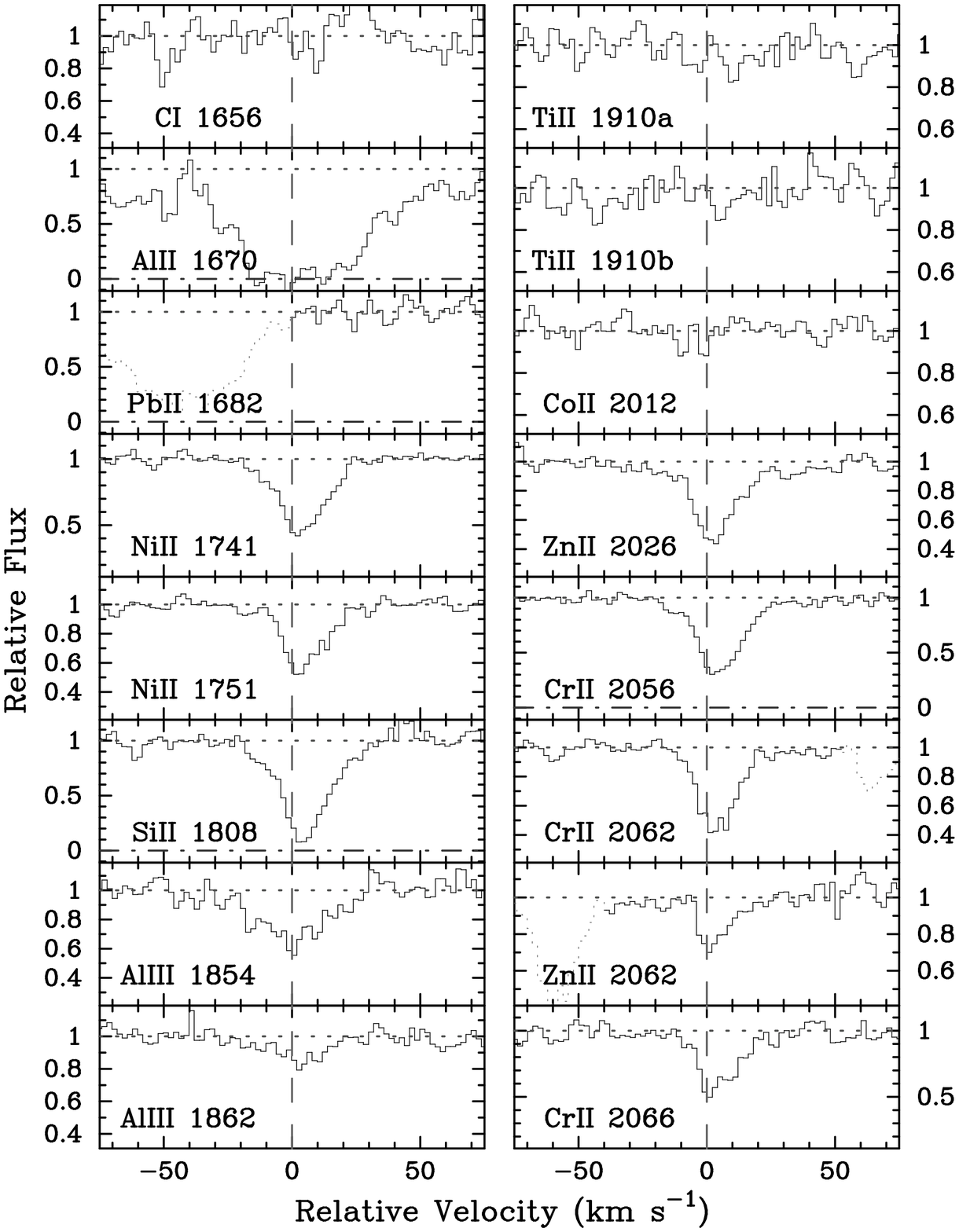}
\caption{Velocity plot of the metal lines in the $z=1.479$ DLA toward
Q0933+733. The velocity zeropoint was chosen arbitrarily to match the
peak optical depth of the \ion{Cr}{2} $\lambda$2066 profile.}
\label{fig:0933mtl}
\end{figure}

\begin{figure*}
\epsscale{0.85}
\plotone{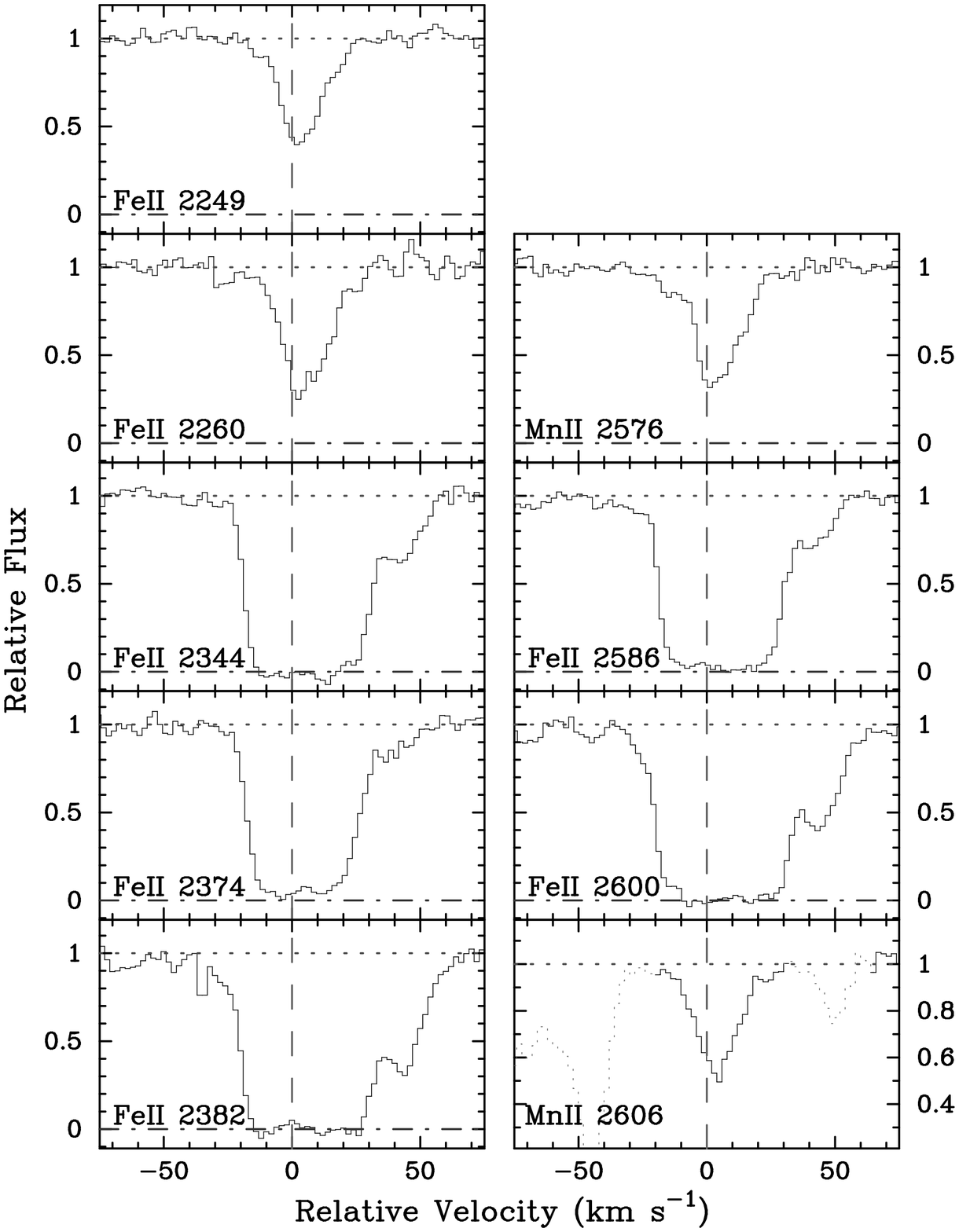}
\end{figure*}

\begin{figure}
\centerline{\includegraphics[scale=0.7,angle=90]{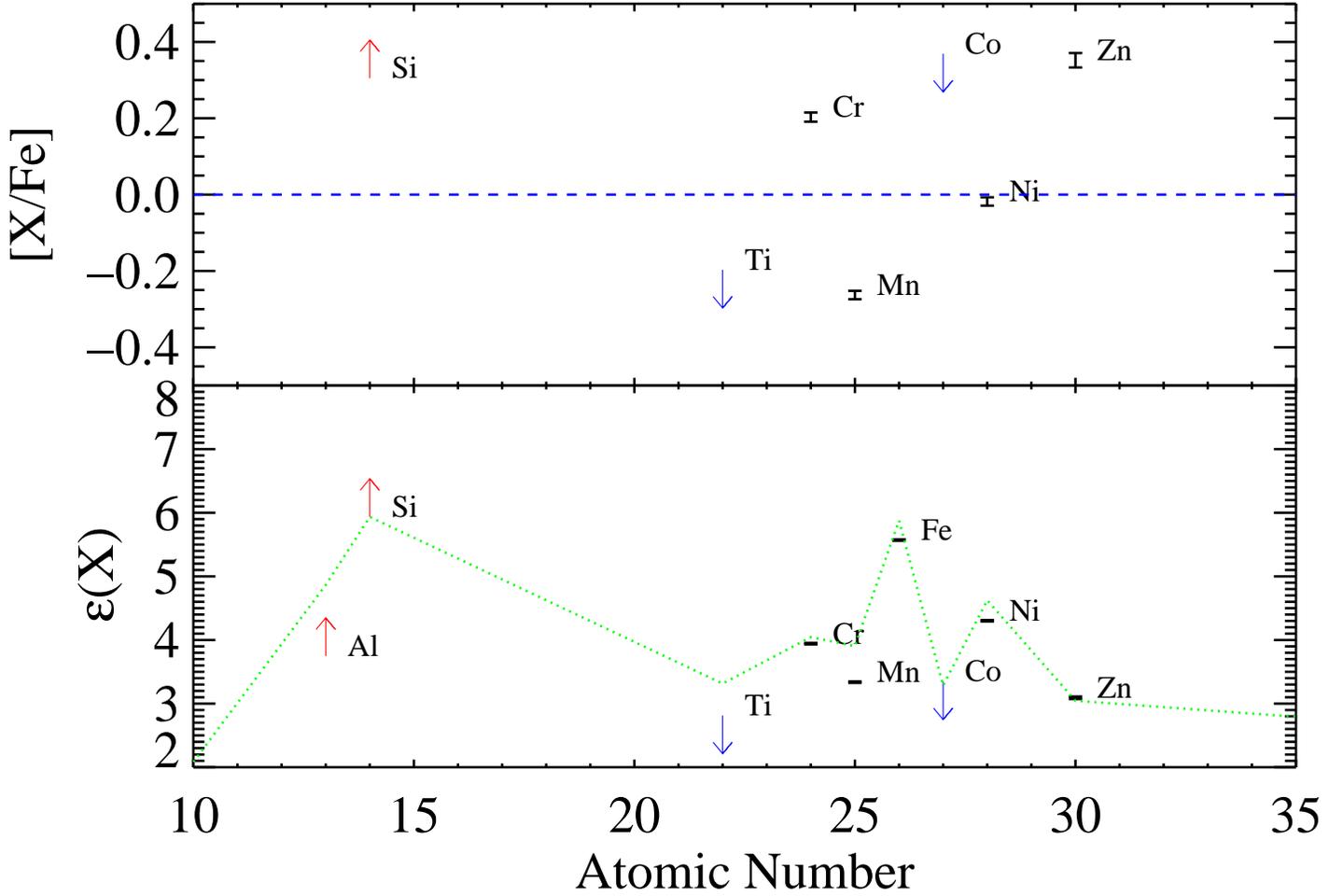}}
\caption{Relative abundances in the $z=1.479$ DLA toward Q0933+733.
The top panel shows the abundances relative to Fe and the bottom panel
gives the absolute abundances on a logarithmic scale with hydrogen
at $\varepsilon(H)=12.0$. The dotted line traces the solar abundance pattern scaled
to match the Zn abundance in the $z=1.479$ DLA. 
These are gas phase abundances uncorrected for depletion.
}
\label{fig:0933rel}
\end{figure}

\begin{figure}
\epsscale{0.85}
\plotone{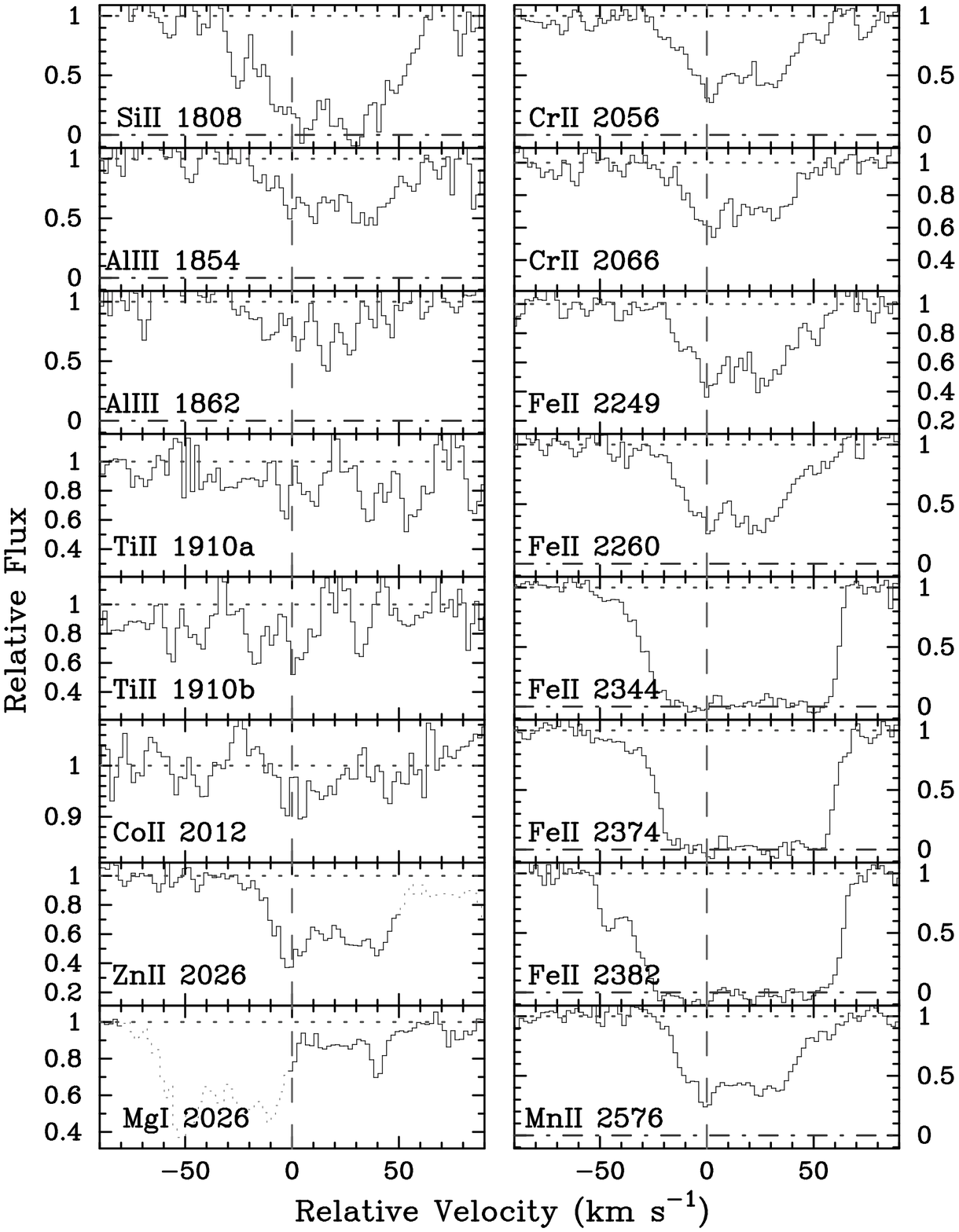}
\caption{Velocity plot of the metal lines in the $z=1.233$ DLA toward
Q0948+433. The velocity zeropoint was chosen arbitrarily to match the
peak optical depth of the \ion{Mn}{2} $\lambda$2576 profile.}
\label{fig:0948mtl}
\end{figure}

\begin{figure*}
\plotone{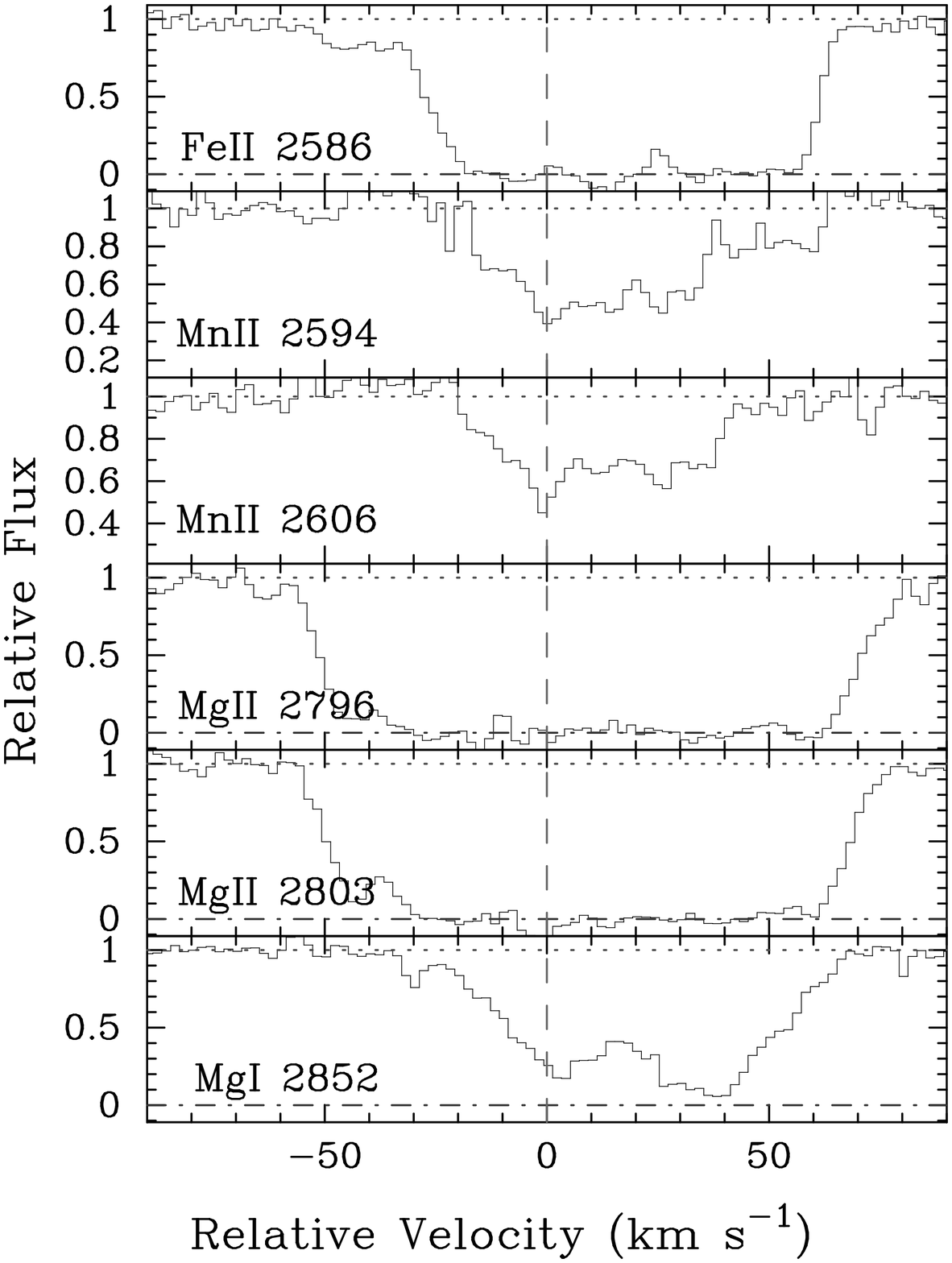}
\end{figure*}

\begin{figure}
\centerline{\includegraphics[scale=0.7,angle=90]{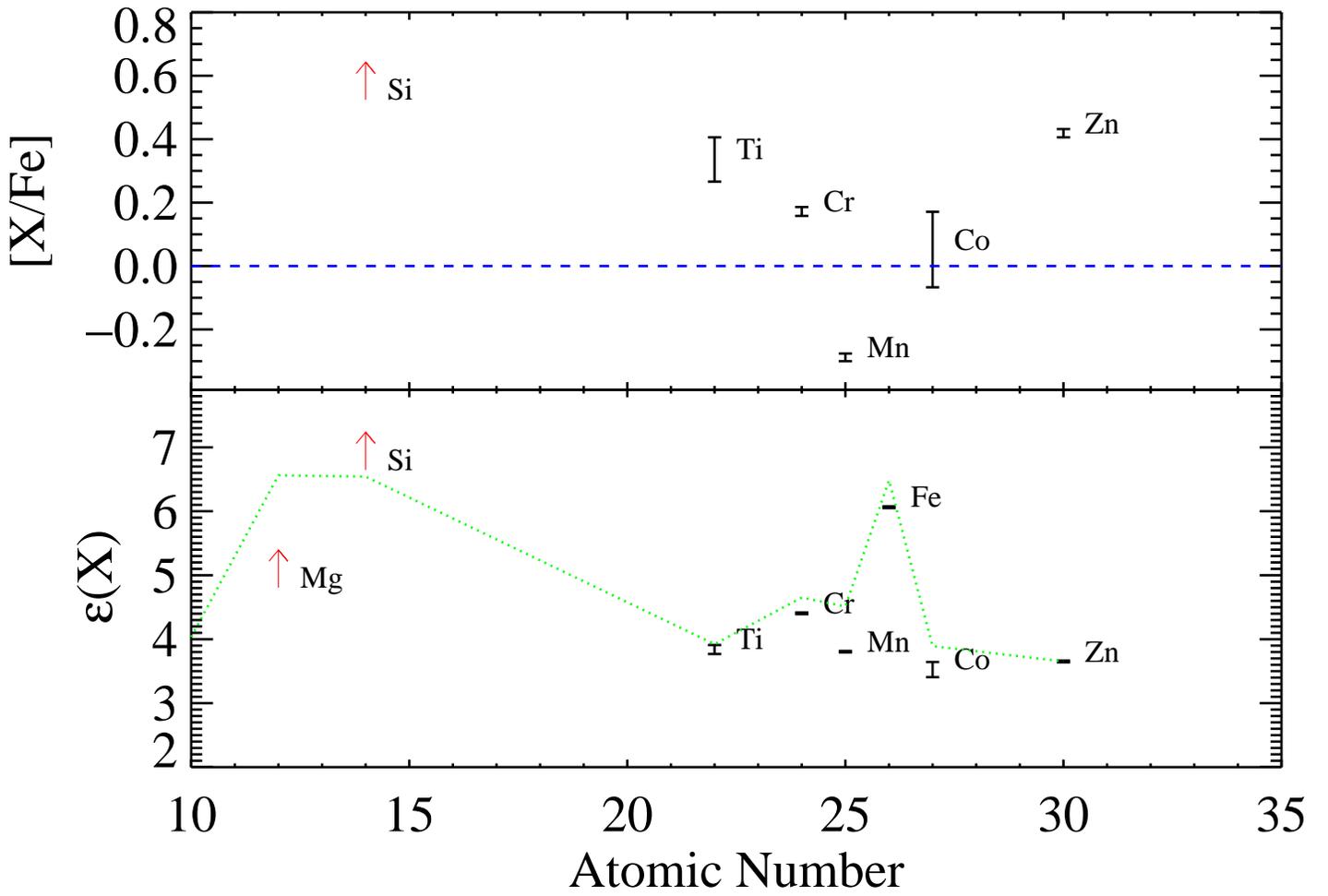}}
\caption{Same as for Figure 3, but for the $z=1.233$ DLA toward Q0948+433.}
\label{fig:0948rel}
\end{figure}

\begin{figure}
\centerline{\includegraphics[scale=0.7,angle=-90]{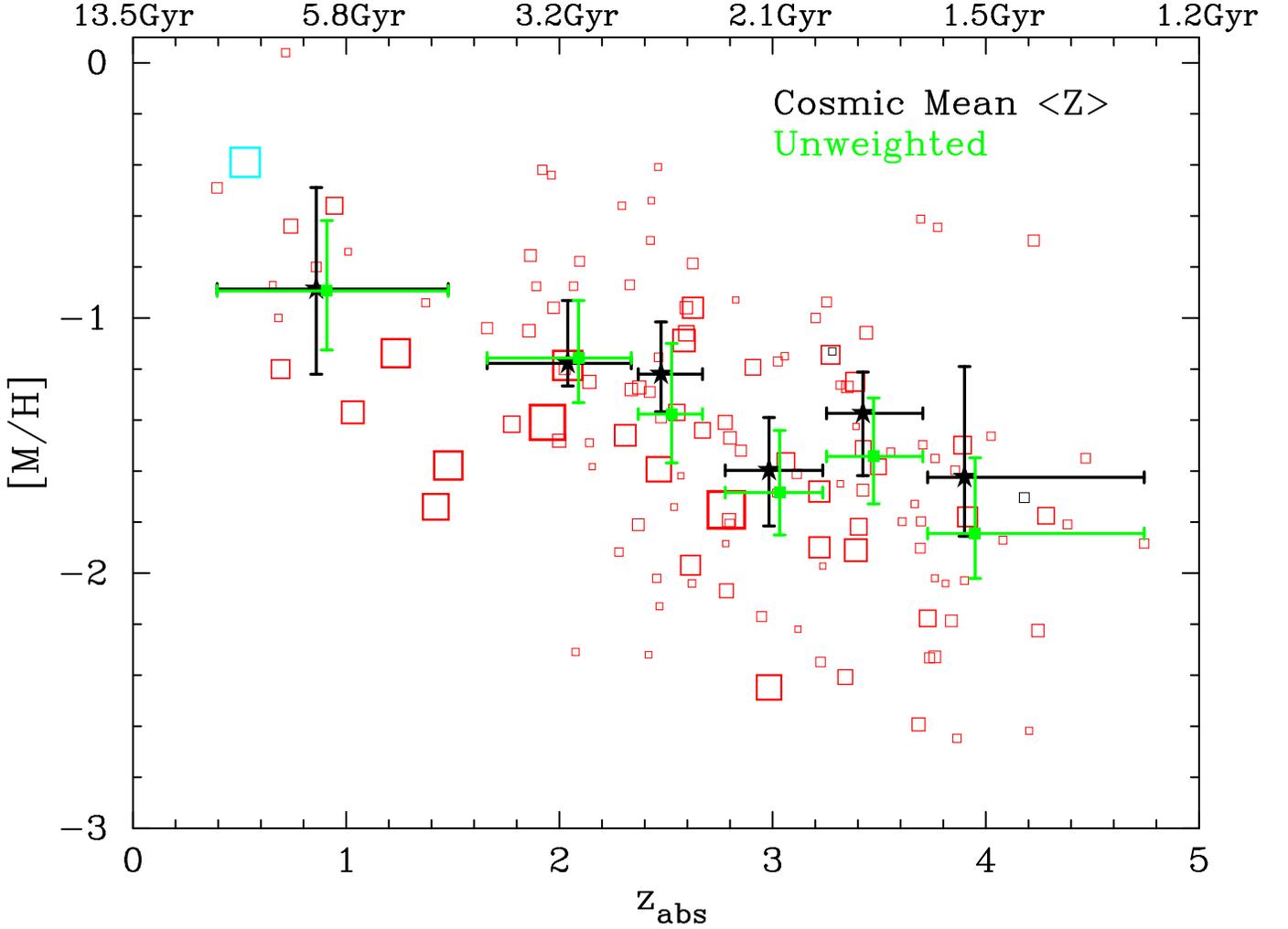}}
\caption{Metallicity [M/H] vs. $z_{abs}$ for DLAs plotted as open squares where the
area of each square is proportional to $N(HI)$.  The X-ray measurement of [M/H] for
the DLA toward Q0235+164 is plotted as the blue open square. All other measurements are
derived from UV or optical quasar spectroscopy. DLAs with larger values of $N(HI)$
dominate the $N(HI)$-weighted mean metallicity. This cosmic mean metallicity, 
$\left<Z\right>$, is plotted for six bins and shown as black filled stars.
Green filled squares are unweighted mean metallicities. Errors are 95\% confidence limits
determined using a bootstrap technique. The lowest redshift value of $\left<Z\right>$
is now more secure since the number of available [M/H] measurements has nearly doubled
since the Prochaska et al. (2003b) analysis. The higher redshift points are adopted from 
Prochaska et al. (2003b). The best-fit slope to the $\left<Z\right>$ vs. $z_{abs}$ data points,
assuming a linear solution, is $m=-0.26 \pm 0.06$, confirming the trend of 
increasing metallicity with cosmic time.}
\end{figure}

\end{document}